# Deterministic Path Search Algorithm on Free-Energy Landscape using Random Grids

Tetsuro Nagai* and Koji Yoshida

*Department of Chemistry, Faculty of Science, Fukuoka University, Fukuoka 814–0180, Japan*

Given a multidimensional free-energy or potential-energy landscape, finding reaction paths that connect an initial (or reactant) state and a final (or product) state is important for biophysics and materials science. The likelihood of a path can be evaluated using an objective function, and the most likely reaction path can be found by optimizing its objective function. However, nonlinear optimization on a complex free-energy or potential-energy landscape may lead to suboptimal solutions. In this study, this drawback is avoided using deterministic path-finding methods such as Dijkstra's algorithm on a graph by assigning grids on the coordinate system to graph nodes and relating the objective function of the path to the edge cost between the nodes. Furthermore, the use of random grids is proposed because they more accurately represent paths than regular grids. As a demonstration, the proposed method is successfully applied to find the minimum resistance path on a three-hole potential model, demonstrating that the proposed method is promising.

*E-mail: tnagai@fukuoka-u.ac.jp

# 1. INTRODUCTION

To understand complex chemical reactions, it is necessary to find a path connecting an initial (or reactant) state and a final (or product) state on a multidimensional free-energy landscape or potential-energy landscape. Whereas for isolated small molecules we may able to deal with the potential energy landscape of all degrees of freedom, for complex systems we need to consider the free energy lansdacpe[1] (also referred to as the potential of mean force and free energy along a reaction coordinate, among other terms), which can be evaluated using molecular dynamics methods.[2-11] Finding a path on these landscape is important analysis because it explains the reaction mechanism.

Even if the free-energy or potential-energy landscape is given, it is far from trivial to identify the optimal path among countless paths in a multidimensional space. To find the optimal path numerically, Karplus[12] and Pratt[13] proposed considering the extreme value problem of the objective function of a path. The most plausible path can be found by first estimating an appropriate path and then numerically optimizing it with respect to the objective function. Another method is the nudged elastic band method[14], to which the string method[15-18] is closely related. The optimal path is found using these methods by optimizing the elastic band that connects the reactant and product states. However, because these methods involve the execution of nonlinear optimization in a complex free-energy or potential-energy landscape, the optimized path may converge to a suboptimal solution. Furthermore, some parameters used in these methods need to be optimally chosen for the respective problems.[19]

These drawbacks can be avoided by adopting deterministic methods to identify optimal paths, such as Dijkstra's algorithm.[20] Dijkstra's algorithm can identify the shortest paths between nodes in a weighted graph. To use this method to find the optimal path on a free-energy or potential-energy landscape, the nodes of the graph must be assigned to the coordinate system of the free-energy or potential-energy landscape. The cost reflecting these landscape must be assigned to the edges of these nodes. Thus, the path-finding problem on these landscapes is reduced to a path-finding problem on weighted graphs, which can be solved deterministically without numerical optimization. A closely related algorithm is implemented as a general-purpose program[21] to find an optimal path on an energy surface. Furthermore, the minimum elevation path between Geneva (in Switzerland) and Turin (in Italy) was also found by the program.[21] However, when the nodes are assigned to the regular grids of coordinate systems, the path length that moves diagonally to the regular grid is overestimated. This overestimation depends on the angle to the grids, resulting in a systematic bias in the resulting paths.

In this study, we propose the use of random grids to eliminate the angle-dependent bias that is inevitable with regular grids. In other words, a path is represented by connecting random grids on the coordinate system.

Using random grids in conjunction with a deterministic path-finding algorithm, namely, Dijkstra's algorithm, we propose a new method for deterministically finding an optimal path with reasonable accuracy. the proposed method is compatible with the objective function that is formulated as the line integral along a path. In this study, we first confirm that the path length represented on random grids is reasonable. We then demonstrate the proposed method by using it to find the minimum resistance path.[22]

The objective function to find the minimum resistance was proposed by Berkowitz et al.[22] and used by Huo and Straub in the MaxFlux method.[23] In this study, we focus on overdamped diffusive dynamics to define the objective function. We consider the energy landscape $U(\boldsymbol{r})$ in accordance with previous studies.[22, 23] However, it is straightforward to consider the free-energy landscape instead. When we consider overdamped dynamics on free-energy landscape $F(\boldsymbol{r})$, we just substitute $U(\boldsymbol{r})$ for $F(\boldsymbol{r})$. The probability distribution is well described using the Smoluchowski equation,

$$\frac{\partial p(\boldsymbol{r},t)}{\partial t} = -\nabla \cdot \boldsymbol{j} \tag{1}$$

where the flux is given by

$$\boldsymbol{j} = -e^{-\beta U(\boldsymbol{r})}\boldsymbol{D}(\boldsymbol{r}) \cdot \nabla\left[p(\boldsymbol{r},t)e^{\beta U(\boldsymbol{r})}\right] \tag{2}$$

where $\beta$ and $\boldsymbol{D}(\boldsymbol{r})$ denote the inverse temperature and diffusion tensor, respectively. Here, we assume that the tensor is isotropic for simplicity, i.e., $\boldsymbol{D}(\boldsymbol{r}) = D(\boldsymbol{r})$. We then assume that $\boldsymbol{j}$ is constant along the reacting path from the initial (reactant) point $\boldsymbol{x}_\mathrm{R}$ to the final (product) point $\boldsymbol{x}_\mathrm{P}$, which is equivalent to assuming that all reactive trajectories are nonintersecting. Then, it can be shown that the flux is maximum if the following line integral

$$R = \int_{x_\text{R}}^{x_\text{P}} \frac{\exp(+\beta U(\boldsymbol{r}))}{D(\boldsymbol{r})} \, dl(\boldsymbol{r}). \tag{3}$$

is minimum. Thus, this line integral can be considered the resistance and used as an objective function to define an optimal path. When $D(\boldsymbol{r})$ is constant, instead of Eq. 3, we can use

$$R = \int_{x_\text{R}}^{x_\text{P}} \exp(+\beta U(\boldsymbol{r})) \, dl(\boldsymbol{r}), \tag{4}$$

as an objective function.

The advantage of this method is that as the resistance is defined on the basis of the reaction rate at finite temperatures, it can address the temperature dependence of the reaction paths. Inversely, at the limit of $\beta \to \infty$, the minimum resistance path converges to the minimum energy path, yielding the exact transition state.[24] Another advantage of this approach is that the position-dependent diffusion constant can be easily incorporated, as mentioned above. The position-dependent diffusion constant can be evaluated through molecular dynamics calculations using advanced techniques. [25-31] Thus, large-scale molecular dynamics simulations enable the calculation of not only the free-energy landscape[32] but also the position-dependent diffusion constant[33] in a three-dimensional (3D) space. Therefore, an efficient path-finding method that can incorporate the position-dependent diffusion constant is indispensable.

This paper is organized as follows: Section II describes the deterministic path-finding method on random grids and presents the numerical details of the application considered in this study. Section III presents the results and discussion. We show that the proposed method has no systematic angle-dependent errors and that the path length is reproduced with satisfactory accuracy. The validity of the proposed method is demonstrated by applying it to the minimum resistance path problem. In the last section, we present our conclusions.

## 2. METHODS

### 2.1. Overview of deterministic path optimization on random grids

In the proposed method, by assigning nodes on a graph to grids on a coordinate system of the free-energy or potential-energy landscape, we present a path on the coordinate system as a path on the graph. We use grids that are randomly placed on the coordinate system rather than regular grids. We assign a cost reflecting the free-energy or potential-energy landscape to the edges of the graph. For any pair of the initial and final nodes, we can find the optimal path in which the sum of the costs along the path is the minimum using Dijkstra's algorithm. In this section, we first review Dijkstra's algorithm, followed by a description of how to construct random grids and assign the cost to each edge. Using a certain form of the cost of the edge, the total cost along a path approximates the line integral of an arbitrary function along the path. Therefore, our method can find a path that minimizes the line integral of the function along a path.

## 2.2. Dijkstra's algorithm

To briefly describe Dijkstra's algorithm, we consider a graph. As an example, we consider a graph consisting of eight nodes, as shown in Fig 1. The cost of $C_{i,j}^{\text{edge}}$ is assigned to each edge. Here, we write a path using a sequence of node numbers such as $\{i_0, i_1, i_2, \ldots i_N\}$, and the cost of a path is given by the sum of $C_{i,j}^{\text{edge}}$ along the path. Therefore, the total cost can be written as follows:

$$C^{\text{total}}(\{i_0, i_1, i_2, \ldots i_N\}) = \sum_{j=0}^{N-1} C_{i_j, i_{j+1}}^{\text{edge}} \,. \tag{5}$$

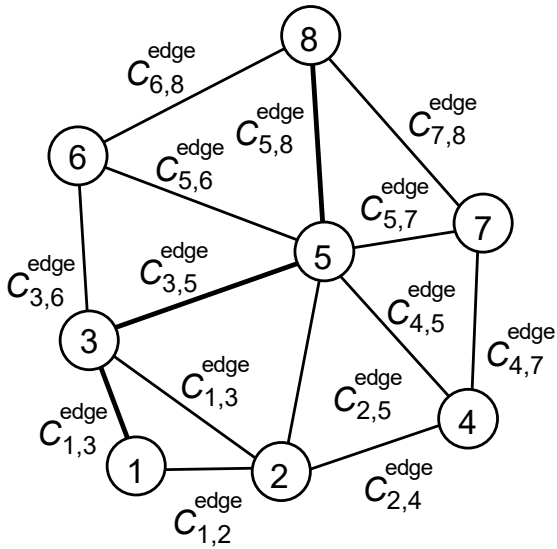

**Fig. 1** Example of the weighted graph consisting of 8 nodes. Each edge assumes the cost of $C_{i,j}^{\text{edge}}$.

For example, the path {1,3,5,8} assumes the cost of $C_{1,3}^{\text{edge}} + C_{3,5}^{\text{edge}} + C_{5,8}^{\text{edge}}$.

Using Dijkstra's algorithm, the path that minimizes $C^{\text{total}}(\{i_0, i_1, i_2, \ldots i_N\})$ can be obtained deterministically for the arbitrary initial node $i_0$ and final node $i_N$. Note that we do not have to specify the number of nodes along a path, $N$, which can be automatically determined. For example, if the path {1,3,5,8} is found to be optimal from node 1 to node 8, any other values of $C^{\text{total}}(\{1, i_1, i_2, \ldots, 8\})$, such as $C_{1,2}^{\text{edge}} + C_{2,5}^{\text{edge}} + C_{5,8}^{\text{edge}}$, $C_{1,3}^{\text{edge}} + C_{3,6}^{\text{edge}} + C_{6,8}^{\text{edge}}$, and $C_{1,2}^{\text{edge}} + C_{2,4}^{\text{edge}} + C_{4,7}^{\text{edge}} + C_{7,8}^{\text{edge}}$, are larger than $C_{1,3}^{\text{edge}} + C_{3,5}^{\text{edge}} + C_{5,8}^{\text{edge}}$. Therefore, we are never trapped in any suboptimal solution.

## 2.3. Correspondence between coordinate system and graph to define paths

In the proposed method, we establish a correspondence between the coordinate system and a graph. We consider randomly and uniformly placed grids $x_1, x_2, \ldots x_L$ on the coordinate system. These random grids $x_1, x_2, \ldots x_L$ are assigned to nodes $1, 2, \ldots, L$ on the graph. A path in the coordinate system is represented by connecting neighboring grids. Thus, the sequence of these grids $\{x_{i_0}, x_{i_1}, \ldots x_{i_N}\}$ is the path on the coordinate system. This sequence corresponds to the sequence of nodes $\{i_0, i_1, i_2, \ldots i_N\}$, which represents the path on the graph.

We then assign a cost of the edge between nodes $i$ and $j$ ($C_{i,j}^{\text{edge}}$) such that the cost reflects the free-energy or potential-energy landscape at $x_i$ and $x_j$ of the coordinate systems. Once $C_{i,j}^{\text{edge}}$ is assigned, the path-finding problem in the continuous coordinates from $x_{i_0} = x_{\text{R}}$ to $x_{i_N} = x_{\text{P}}$ is reduced to a path-finding problem on a graph from node $i_0$ to node $i_N$, which can be solved deterministically using Dijkstra's algorithm.

To find neighboring grids on the coordinate systems, we use Delaunay triangulation[34]. Delaunay triangulation in a plane has no vertex inside the circumscribing circle of any triangle. In the three dimensions, no vertex is enclosed by the circumscribing sphere of any tetrahedron. If an edge between two grids $x_i$ and $x_j$ is connected by Delaunay triangulation, $x_i$ and $x_j$ can be connected to represent a path. Thus, we assign the finite value for the cost $C_{i,j}^{\text{edge}}$ to the edge between two corresponding nodes $i$ and $j$. If an edge between two grids $x_i$ and $x_j$ is not connected by Delaunay triangulation, we do not consider the connection between the two grids. In this case, we assume that there is no edge with a finite cost between nodes $i$ and $j$ on the graph.

For the assignment of $C_{i,j}^{\text{edge}}$, we propose the following approach:

$$C_{i,j}^{\text{edge}} = \frac{f(x_i) + f(x_j)}{2} d_{i,j}, \tag{6}$$

where $d_{i,j}$ denotes the distance between $x_i$ and $x_j$, and $f(x_i)$ can be an arbitrary function on the corresponding coordinate system. For example, we can use the free-energy landscape. The form of Eq. 6 is useful as the sum of $C_{i,j}^{\text{edge}}$ approximate the line integral of $f(x)$.

$$C^{\text{total}}(\{i_0, i_1, i_2, \ldots i_N\}) \approx \int_{x_R}^{x_P} f(x) dl(x) \tag{7}$$

As such, by finding a path that minimizes $C^{\text{total}}(\{i_0, i_1, i_2, \ldots i_N\})$, we can identify a path that approximately minimizes the line integral $\int_{x_R}^{x_P} f(x) dl(x)$. This correspondence is helpful in connecting the proposed method to theoretical frameworks. For example, to find the minimum resistance path with respect to Eq. 3, we set $f(x) = \exp(\beta U(x))/D(x)$.

The use of Dijkstra's algorithm imposes the condition that $C_{j,j+1}^{\text{edge}}$ must be positive, i.e., $C_{j,j+1}^{\text{edge}} \geq 0$. Presumably, this condition is not severe in most applications. When we use the free-energy landscape for $f(x)$, we can change the reference state such that $f(x) \geq 0$. For the minimum resistance problem, $C_{j,j+1}^{\text{edge}} \geq 0$ holds by definition.

The numerical details of this study are as follows. Random grids were generated using the pseudorandom generator Mersenne Twister.[35] For Delaunay triangulation, we used Qhull's package.[36] Dijkstra's algorithm was implemented in C++.

### 2.4. Application to path of minimum distance

To verify the ability to represent distances on random grids, we obtained an optimal path for

$$C_{j,j+1}^{\text{edge}} = d_{i_j, i_{j+1}}. \tag{8}$$

In the continuous limit, the path is the line segment that interconnects $x_R$ and $x_P$. The value of $C^{\text{total}}(\{i_0, i_1, i_2, \ldots i_N\})$ is the distance between $\boldsymbol{x}_R$ and $\boldsymbol{x}_P$, which is represented by connecting neighboring random grids. Therefore, this is a good test case for studying the error of the distance in the proposed method.

In this application, the initial point was set to the origin of the two-dimensional (2D) space, i.e., $\boldsymbol{x}_R = (0,0)$. The final point was set to $\boldsymbol{x}_{P,k} = (\cos\theta_k, \sin\theta_k)$, where $\theta_k = \frac{2\pi}{60}k$ for $k = 0, 1, \ldots, 59$. The average number of random grids per unit area was varied such that $\rho_{\text{grids}} = 100$, 400, 2500, 10000, 40000, 250000, and 1000000. The actual distance is unity; thus, the difference between $C^{\text{total}}(\{i_0, i_1, i_2, \ldots i_N\}; \theta_k, \rho_{\text{grids}})$ and unity is an indicator of the accuracy of the approximation of path length. Therefore, this is a good test for checking the systematic error depending on the direction of the path and grid density. Another test in 3D space is presented in supplemental materials[37].

## 2.5. Application to the minimum resistance path in three-hole potential model

In this study, we examin the 2D three-hole potential studied by Huo and Straub.[23] The potential form is given by

$$U_{\text{TH}}(x,y) = -3.0\, e^{-x^2}\left[e^{-\left(y-\frac{5}{3}\right)^2} - e^{-\left(y-\frac{1}{3}\right)^2}\right] - 5.0\, e^{-y^2}\left[e^{-(x-1)^2} + e^{-(x+1)^2}\right]. \tag{9}$$

This model has three potential energy minima (holes), as shown in Fig. 2. The reactant and product states are the left and right minima, respectively, on the front side. The third minimum is behind the two holes. Therefore, two paths are possible. In the first path, two minima are directly connected through the saddle point between the two minima. The second path passes through the third minimum, surmounting the two saddle points. The saddle point in the first path is higher than that in the second path; however, the first path is shorter.

To obtain the minimum resistance path according to the MaxFlux method[23], by substituting $U(\boldsymbol{x}) = U_{\text{TH}}(x,y)$ in Eq. 4, we need to find the path that minimizes the resistance,

$$R = \int_{\boldsymbol{x}_R}^{\boldsymbol{x}_P} \exp\left(+\beta U_{\text{TH}}\left(x_{i_j}, y_{i_j}\right)\right) dl(\boldsymbol{x}). \tag{10}$$

At higher temperatures, the height of the barrier becomes less important than at lower temperatures and the optimal path is the shorter first path. At lower temperatures, barrier crossing is more challenging. Thus, the second path is preferred because although it is longer, the saddle points are lower than that for the first path.

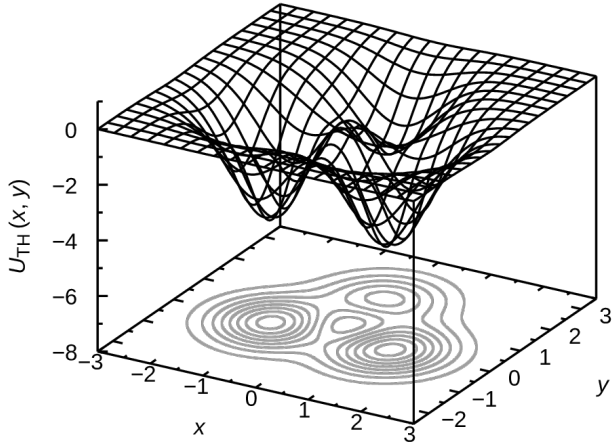

**Fig 2.** Shape of the three-hole model used in this study.

### 2.5.1. Application of the proposed method to find the path of minimum resistance

In the application of the proposed method, by substituting $f(x) = \exp(+\beta U_{\text{TH}}(x,y))$ to Eq. 6, the cost of the edge between nodes $i$ and $j$ is given by

$$C_{i,j}^{\text{edge}} = \frac{\exp(+\beta U_{\text{TH}}(x_i, y_i)) + \exp(+\beta U_{\text{TH}}(x_j, y_j))}{2} d_{i,j}, \tag{11}$$

where $\beta$ denotes the inverse temperature. Therefore, the line integral representing the resistance, Eq. 10, was approximately minimized. The grid density was $\rho_{\text{grids}} = 1000000$. Dijkstra's algorithm, coded as a serial program, took less than 5 min to perform using an Intel(R) Xeon(R) Gold 6136 CPU at 3.00 GHz.

Technically speaking, this demonstration is performed on the model potential-energy landscape. However, by using a free energy landscape $F(r)$ instead of $U(r)$ we can find a path on the free energy landscape.

### 2.5.2. Numerical details of conventional methods

A method based on square grids is implemented for comparison with the proposed method. Square grids with an interval of $\Delta x = 0.001$ are considered. Paths connecting the nearest neighboring grids are considered, and the cost function is

$$C_{\text{square}}^{\text{total}}(\{i_0, i_1, i_2, \ldots i_N\}) = \sum_{j=0}^{N-1} \frac{\exp\left(+\beta U_{\text{TH}}\left(x_{i_j}, y_{i_j}\right)\right) + \exp\left(+\beta U_{\text{TH}}\left(x_{i_{j+1}}, y_{i_{j+1}}\right)\right)}{2} \Delta x. \quad (12)$$

The path with the smallest $C_{\text{square}}^{\text{total}}(\{i_0, i_1, i_2, \ldots i_N\})$ is found using Dijkstra's algorithm.

We also reproduce the work of Huo and Straub.[23)] Therefore, we minimize the same objective function as that in Ref[23)]. The discretized version of Eq. 4, namely Eq. 7 in Ref [23)]

$$\mathcal{R} = \sum_{k=0}^{N-1} e^{\beta U(r_k)} |r_{k+1} - r_k| \quad (13)$$

is the objective function. Here, $N + 1$ particles represent a path on the potential-energy landscape, and $r_k$ denotes the position of the $k$th particle. Unlike the proposed method, $r_k$ is a continuous variable and is subject to numerical optimization. Following the literature[23)], some other auxiliary terms are added to make the particles a self-avoiding chain separated at a constant interval, which are given by

$$\mathcal{C}_A = \kappa \sum_{k=1}^{N} [(r_k - r_{k-1})^2 - d_{\text{ave}}^2]^2 \quad (14)$$

and

$$\mathcal{C}_R = \frac{\rho}{\lambda} \sum_{j > k+1} \exp\left[-\lambda (r_j - r_k)^2 / \langle d \rangle^2\right] \quad (15)$$

where $d_{\text{ave}}^2 = \sum_{k=1}^{N}(r_k - r_{k-1})^2/N$ and $\langle d \rangle = \sum_{k=1}^{N}|r_k - r_{k-1}|/N$. Here, $\kappa$, $\rho$, and $\lambda$ denote parameters that need to be tuned for the respective problems. In this study, we use the following

conditions: $N = 25$, $\lambda = 2.0$, $\rho = 0.05$. We set to $\kappa = 5.0$ for $\beta = 1.0, 2.0$, and 3.3 and $\kappa = 0.5$ for $\beta = 4.0$. The The optimization is performed using an annealing procedure in accordance with Ref[23]. In our implementation, we sample $r_k$ with respect to the Boltzmann distribution of $\exp[-(\mathcal{R} + \mathcal{C}_A + \mathcal{C}_B)/T_a]$ using the Monte Carlo technique,[38] where $T_a$ denotes a fictitious annealing temperature. Initially, $T_a$ is increased such that various possible reactive paths are sampled. Then, $T_a$ is gradually decreased to minimize $\mathcal{R} + \mathcal{C}_A + \mathcal{C}_B$, and the self-avoiding chain will settle down to an estimate of the reaction pathway with the maximum flux.

The MaxFlux method and our method for finding the minimum resistance path are similar in that both methods aim to find the path that minimizes Eq. 4 (or more generally Eq. 3). The difference is primarily in the minimization process. In the former, the path is represented using connected particles that move continuously on the original coordinate system. The weight of sampling is designed such that these particles find suitable places to minimize Eq. 4 through the annealing process. In the latter, a path is represented by connecting predefined random grids and a cost is assigned to each edge such that their sum corresponds to Eq. 4. By reducing the problem to a path-finding problem on a weighted graph, the optimal path can be obtained deterministically without obtaining a suboptimal solution due to the annealing scheme.

## 3. RESULTS AND DISCUSSION

### 3.1. Application to path of minimum distance

Figure 3 shows the optimal paths from the origin to each point on the unit circle obtained using the proposed method on random grids. The obtained paths were approximately straight toward each final point despite the paths being represented on random grids. Figure 4 shows the angular dependence of the path length $C^{\text{total}}_{\text{random grids}}(\theta, \rho_{\text{grids}}) = \sum_{j=0}^{N-1} d_{i_j, i_{j+1}}$. Note that the path length on square grids can be obtained from a trivial calculation:

$$C^{\text{total}}_{\text{square grids}}(\theta) \simeq |\cos \theta| + |\sin \theta|. \tag{16}$$

The path length on the triangle grids is given by

$$C^{\text{total}}_{\text{triangle grids}}(\theta) \simeq \cos \theta' + \frac{1}{\sqrt{3}} \sin \theta', \tag{17}$$

where $\theta' = \theta - \left\lfloor \frac{\theta}{\frac{\pi}{3}} \right\rfloor \frac{\pi}{3}$ and $\lfloor \cdot \rfloor$ represents the floor function. When we use regular grids, irrespective of whether they are square or triangle in shape, the angle-dependent systematic error is large. This error over-penalizes movements in directions that are not along the grid. This angle-dependent error cannot be eliminated by decreasing the grid spacing and is thus inevitable. In contrast, when we use random grids, $C_{\text{random grids}}^{\text{total}}(\theta, \rho_{\text{grids}})$ does not exhibit direction dependence and is only 4% larger than unity. Thus, the proposed method is free from the angle-dependent systematic error, and the constant error is also small. When we consider a line integral as the objective function of a path, the cost can be overestimated in proportion to the estimated path lengths. Therefore, eliminating the large angle-dependent error shown here is essential to effectively describe a path using grids.

Figure 5 shows the accuracy of the distance as a function of grid density. The accuracy is quantified using the following quantities averaged over the angle:

$$\bar{C}_{\text{random grids}}^{\text{total}}(\rho_{\text{grids}}) = \frac{1}{60} \sum_{k=0}^{59} C^{\text{total}}(\theta_k, \rho_{\text{grids}}). \tag{18}$$

Figure 5 indicates that the value of $\bar{C}_{\text{random grid}}^{\text{total}}(\rho_{\text{grids}})$ converges to 1.04. Therefore, the path length is overestimated by only 4% using the proposed method in 2D applications. Similar results are obtained in 3D space, as shown in supplemental materials[37].

The factor of 1.04 is consistent with the factor estimated roughly as follows. When the line segment of unit length is approximated by the sum of the shorter line segments inclined by angle $\pm\theta$, the path length is $\frac{1}{\cos\theta}$. In 2D space, the coordination number of random grids is six; thus, the angle $\theta$ should lie in the range $-\pi/6 < \theta < \pi/6$. Therefore, we average $\frac{1}{\cos\theta}$ for $-\pi/6 < \theta < \pi/6$, which yields

$$\frac{\int_{-\pi/6}^{\pi/6} \frac{d\theta}{\cos\theta}}{\int_{-\pi/6}^{\pi/6} d\theta} = \frac{3\ln 3}{\pi} \approx 1.049. \tag{19}$$

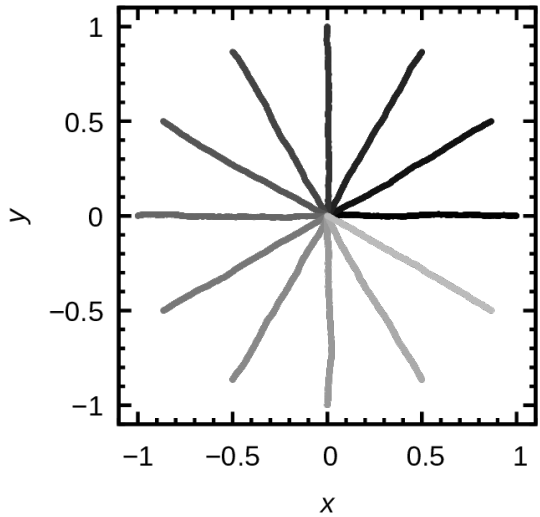

**Fig. 3.** Optimal paths from the origin to $(\cos\theta, \sin\theta)$ for $\theta = \frac{2\pi}{60}i$, where $i = 0, 5, 10, \ldots, 50, 55$. The darkest line represents $i = 0$. The line becomes lighter as $i$ increases, and the lightest line represents $i = 55$. The data shown were obtained for $\rho= 1000000$.

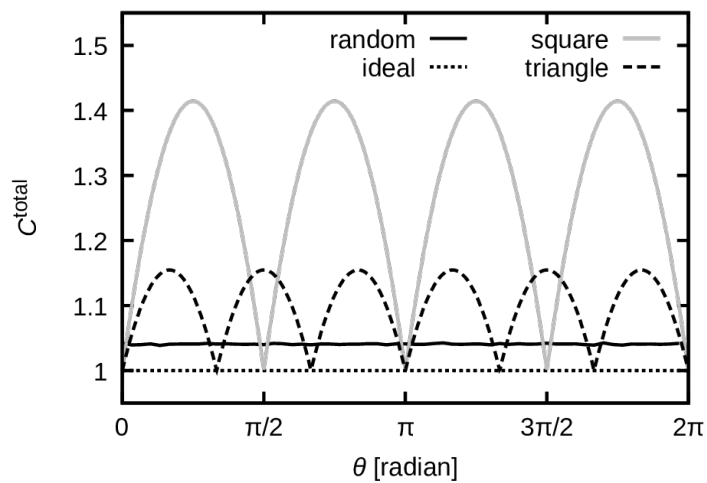

**Fig. 4.** Radial dependence of $C^{total}_{random\ grids}(\theta, \rho_{grids})$, $C^{total}_{square\ grids}(\theta)$, and $C^{total}_{triangle\ grids}(\theta)$ (black, gray, and dashed lines, respectively). For $C^{total}_{random\ grids}(\theta, \rho_{grids})$, data are shown for $\rho_{grids}$ = 1000000. The difference from unity measures the error of the approximation of distance using the grid coordinates.

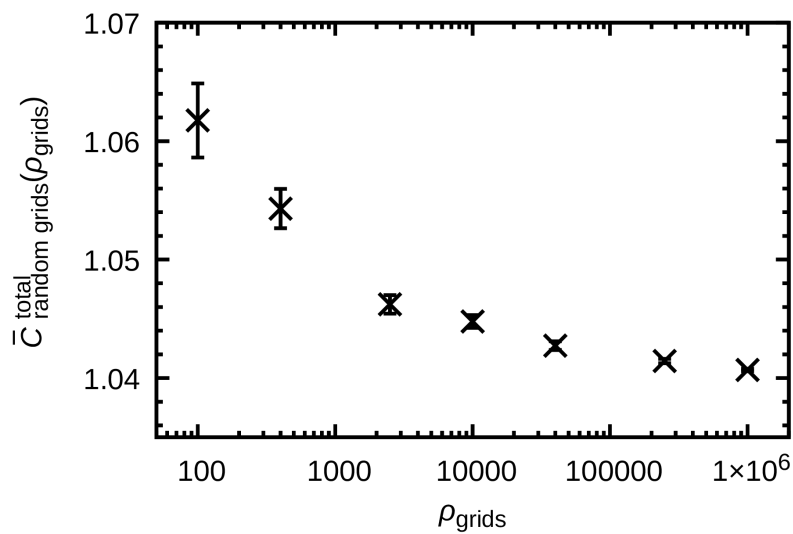

**Fig. 5.** Dependence of $\bar{C}^{total}_{random\ grids}(\rho_{grids})$ on $\rho_{grids}$ in the 2D application. Bars show the standard error of 60 data points.

## 3.2. Application to the minimum resistance path in three-hole potential model

Figure 6 shows the minimum resistance paths obtained using the proposed and conventional methods at each temperature. Although the paths obtained using square grids (gray lines) approximate some of those obtained using the MaxFlux method (particles in the self-avoiding chain are indicated by black dots), stark differences were observed. First, some parts of the path obtained using square grids are unnaturally straight. Second, for $\beta$ = 3.3, the path obtained using the square grids is different from those predicted using our algorithm and the MaxFlux method: the path does not go through the intermediate basin. These artifacts in the square grids stem from the over-penalty of the paths that are oblique to the grids, as shown in Section 3.1. In contrast, at all temperatures, the paths obtained using the proposed method (red lines) and the MaxFlux method (black dots) are similar. The proposed method using random grids succeeds in selecting the same path as that selected using the MaxFlux method even at $\beta$ = 3.3. Thus, our method is rigorous.

The major difference between the MaxFlux and proposed methods is that the former requires nonlinear optimization in a continuous space and may not be able to find the correct path if the choice of parameters or initial conditions is poor. As discussed in Ref [19], this type of method requires careful parameter adjustments. The proposed method is a deterministic algorithm for identifying a path on a graph, where we can find the globally optimal path up to the resolution of random grids.

The only parameter of the proposed method is the grid density determining the resolution of random grids. To examine the effect of $\rho_{\text{grids}}$ on the found path, we show the optimal paths obtained for lower grid densities $\rho_{\text{grids}} = 10000$ in Fig. 7 (data with $\rho_{\text{grids}} = 40000$ are also shown in Fig S2 in supplemental materials[37]). Even with this significantly low grid density, although slightly off due to the deteriorated resolution, the paths found using our method are consistent with those obtained using the MaxFlux method. Therefore, the proposed method is robust to the parameter and useful for efficiently finding an optimal path.

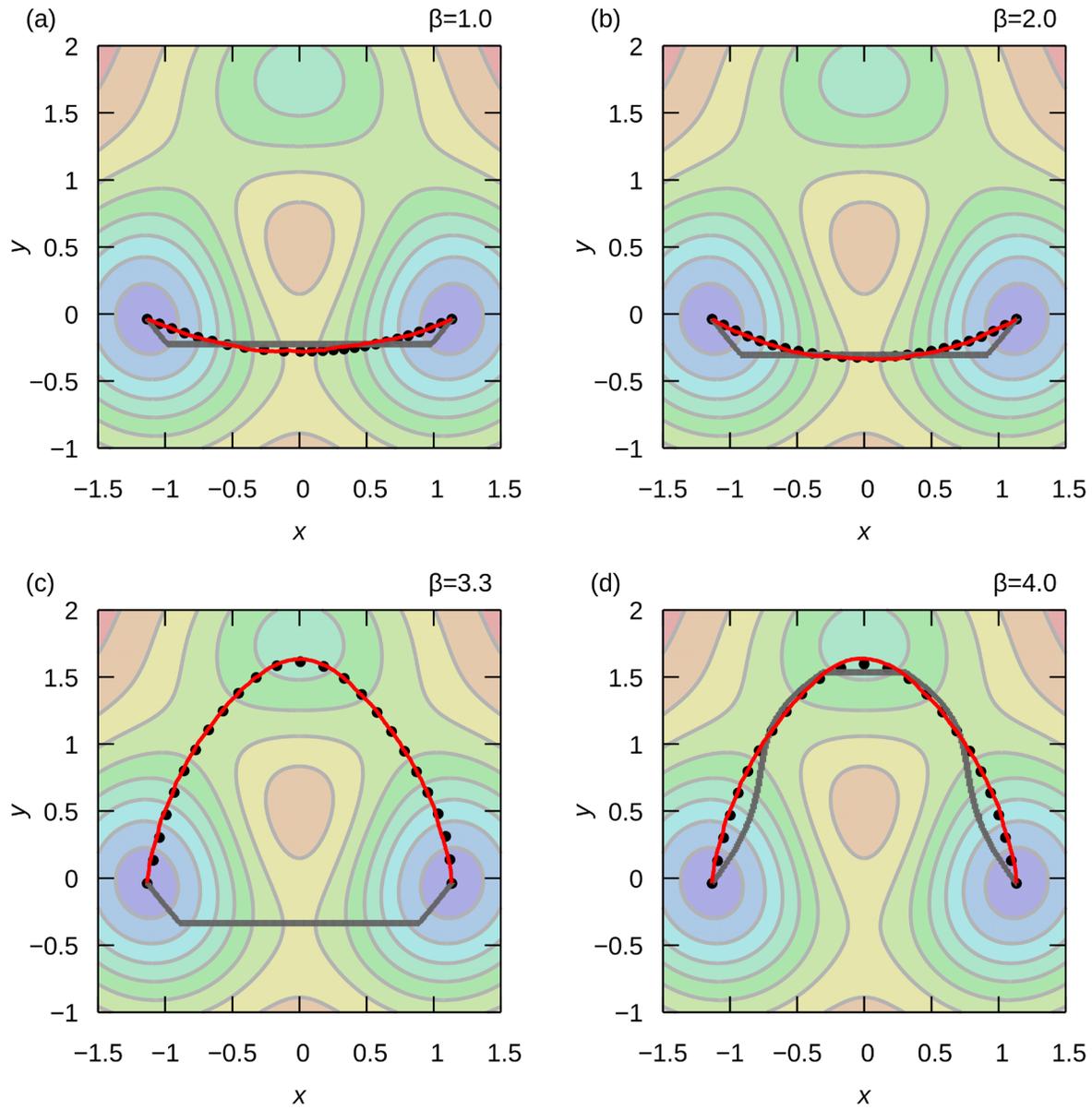

**Fig. 6.** (Color) Comparison of paths obtained using the proposed method (red lines), Dijkstra's algorithm in square grids (gray lines), and the MaxFlux method (particles are shown by black circles) for $\beta =$ (a) 1.0, (b) 2.0, (c) 3.3, and (d) 4.0. For the proposed method, the grid density is set to $\rho_{\text{grids}} = 1000000$. The contours with the heat map represent the values of the potential energy $U_{\text{TH}}(x, y)$.

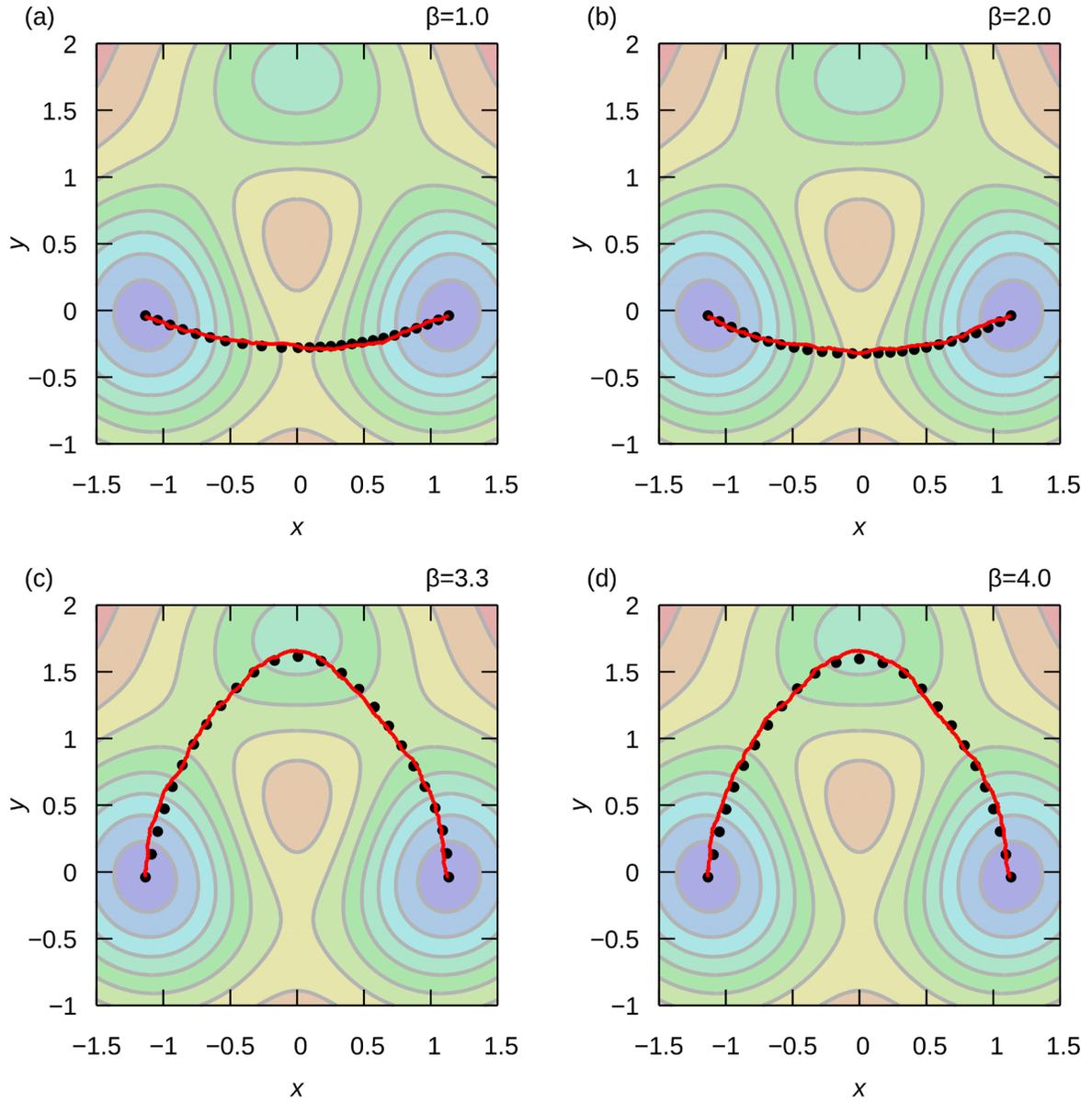

**Fig. 7.** (Color online) Comparison of paths obtained using the proposed method (red lines) and the MaxFlux method (particles are shown by black circles) for $\beta =$ (a) 1.0, (b) 2.0, (c) 3.3, and (d) 4.0. For the proposed method, the grid density is set to $\rho_{\text{grids}} = 10000$. The contours with the heat map represent the values of the potential energy $U_{\text{TH}}(x, y)$.

## 4. CONCLUSIONS

In this study, we proposed a path-finding method based on Dijkstra's algorithm on randomly arranged grids. By imposing the correspondence between a coordinate system and a graph, we can reduce the path-finding problem in the continuous coordinates to that on the graph. Thus, we can use a deterministic method such as Dijkstra's algorithm to find an optimal path without obtaining a suboptimal solution.

We first demonstrated the superiority of random grids over regular grids in an application to find the minimum distance path. Although the path length on regular grids involves a large angle-dependent systematic error from 0% to 41% overestimation, the path length on random grids is accurate with only a small constant error of 4%. When we consider a line integral as an objective function of the path, the cost can be overestimated in proportion to the path length. Therefore, a minor constant error is significantly preferred to an oscillating error with a large amplitude.

In an application of finding the minimum resistance path in a three-hole potential model, we showed that the proposed method can find the optimal paths in accordance with the MaxFlux method. Conversely, some paths were not well reproduced using regular grids, because of over-penalty to paths oblique to the grids. These results reinforce the importance of random grids. The only parameter of the proposed method is the grid density. Even with a significantly lower density of random grids, the proposed method finds the correct paths even though they may be slightly off due to the deteriorated resolution. This observation indicates the robustness of the proposed method. Therefore, the proposed method is promising for finding an optimal path on a free-energy landscape.

## ACKNOWLEDGMENTS

The authors are grateful to Dr. Naoyuki Miyashita of Kindai University for the fruitful discussion. The authors also thank Dr. Zhiye Tang of the Institute of Molecular Science for his guidance in implementing Dijkstra's algorithm. This study was supported in part by the Ministry of Education, Culture, Sports, Science and Technology (MEXT) as the Program for Promoting Research on the Supercomputer Fugaku ("Development and application of a multiscale computational method for the study of the mass transport mechanisms in the fuel cell catalyst layer"; Grant Number JPMXP1020230318). The computational resources of the supercomputer Fugaku were provided by the RIKEN Center for Computational Science (Project ID: hp230200, hp240208, hp210206, hp220238, hp230259). Calculations were also performed on supercomputers at the Research Center for Computational Science at Okazaki (Project: 23-IMS-C136, 24-IMS-C065, 23-IMS-C504, 24-IMS-C502). This study was partially supported by funding from Fukuoka University (GW2404, 215008) and the Japan Society for the Promotion of Science (JSPS) KAKENHI (Grant Number JP22K12673).

# Supplemental Material for Deterministic Path Search Algorithm on Free-Energy Landscape using Random Grids

Tetsuro Nagai* and Koji Yoshida

*Department of Chemistry, Faculty of Science, Fukuoka University, Fukuoka 814–0180, Japan*

*E-mail: tnagai@fukuoka-u.ac.jp

# S1 Application to path of minimum distance in three-dimensional space

The check of the ability of reproducing the distance is also performed in 3D space. In this examination, the initial point was set to the origin, i.e., $x_R = (0, 0, 0)$. The final point was set to $x_{P,k} = (1, 0, 0)$, $(0, 1, 0)$, $(0, 0, 1)$, $(1/\sqrt{2}, 1/\sqrt{2}, 0)$, $(0, 1/\sqrt{2}, 1/\sqrt{2})$, $(1/\sqrt{2}, 0, 1/\sqrt{2})$, and $(1/\sqrt{3}, 1/\sqrt{3}, 1/\sqrt{3})$. The actual distance is unity, and thus the difference between $C^{\text{total}}(\{i_0, i_1, i_2, \ldots i_N\}; x_{P,k}, \rho_{\text{grids}})$ and unity indicates the accuracy of the approximation of path length in the 3D space. The average number of random grids per unit volume was varied: $\rho_{\text{grids}} = 1000, 8000, 125000, 1000000$, and $1953125$.

Table S1 shows the approximated distances $C^{\text{total}}_{\text{random grids}}$ and $C^{\text{total}}_{\text{square grids}}$ from the origin to the seven final points on the unit sphere in 3D space. There is no direction-dependent error for $C^{\text{total}}_{\text{random grids}}$ in 3D space, and the overestimation is only 6%. Conversely, when the path is represented on the square grids, the path length $C^{\text{total}}_{\text{square grids}}$ can be as large as ~1.73. Thus, the distance can be overestimated almost twice. This result indicates that random grids can markedly improve the representation of distances in 3D space. To discuss the grid-density dependence of the path lengths obtained with random grids, we plotted

$$\bar{C}^{\text{total}}_{\text{random grids}}(\{i_0, i_1, i_2, \ldots i_N\}, \rho_{\text{grids}}) = \frac{1}{7} \sum_{k=0}^{6} C^{\text{total}}_{\text{random grids}}(\{i_0, i_1, i_2, \ldots i_N\}; x_{P,k}, \rho_{\text{grids}}) \qquad (S1)$$

as a function of $\rho_{\text{grids}}$ in Fig. S1. This graph indicates that the factor converges at approximately 1.06. A factor of 1.06 is consistent with the factor estimated roughly as follows. We averaged $\frac{1}{\cos\theta}$ in the 3D space for $-\pi/6 < \theta < \pi/6$,

$$\frac{\int_{-\pi/6}^{\pi/6} \frac{2\pi \sin\theta \, d\theta}{\cos\theta}}{\int_{-\pi/6}^{\pi/6} 2\pi \sin\theta \, d\theta} = \frac{\ln 2/\sqrt{3}}{1 - \sqrt{3}/2} \approx 1.073. \qquad (S2)$$

**Table S1.** Cost function $C^{\text{total}}$ for the seven values of $x_P$. The difference from unity measures the error of the approximation of distance using the grid coordinates. The data are for $\rho_{\text{grids}} = 1953125$.

| $x_{P,k}$ | $C^{\text{total}}_{\text{random lattice}}$ | $C^{\text{total}}_{\text{square lattice}}$ |
|---|---|---|
| (1, 0, 0) | 1.059 | 1 |
| (0, 1, 0) | 1.059 | 1 |
| (0, 0, 1) | 1.062 | 1 |
| $(1/\sqrt{2}, 1/\sqrt{2}, 0)$ | 1.061 | $\sqrt{2} \approx 1.41$ |
| $(0, 1/\sqrt{2}, 1/\sqrt{2})$ | 1.057 | $\sqrt{2} \approx 1.41$ |
| $(1/\sqrt{2}, 0, 1/\sqrt{2})$ | 1.062 | $\sqrt{2} \approx 1.41$ |
| $(1/\sqrt{3}, 1/\sqrt{3}, 1/\sqrt{3})$ | 1.061 | $\sqrt{3} \approx 1.73$ |

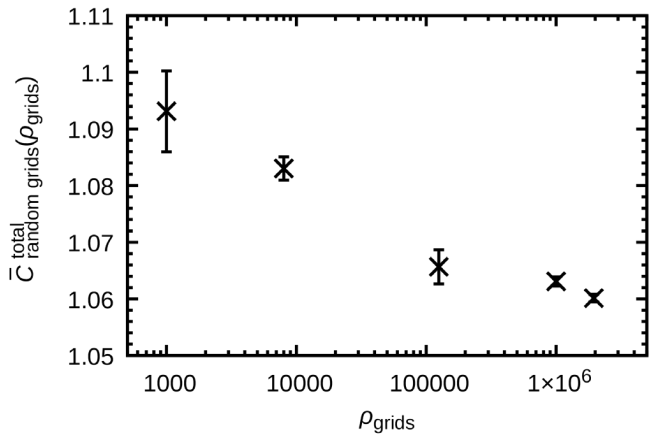

**Fig. S1.** Dependence of $\bar{C}^{\text{total}}_{\text{random grids}}(\{i_0, i_1, i_2, \ldots i_N\}, \rho_{\text{grids}})$ on $\rho_{\text{grids}}$. Bars indicate the standard error evaluated from the seven data points.

## S2 Application to the minimum resistance path in the three-hole potential model

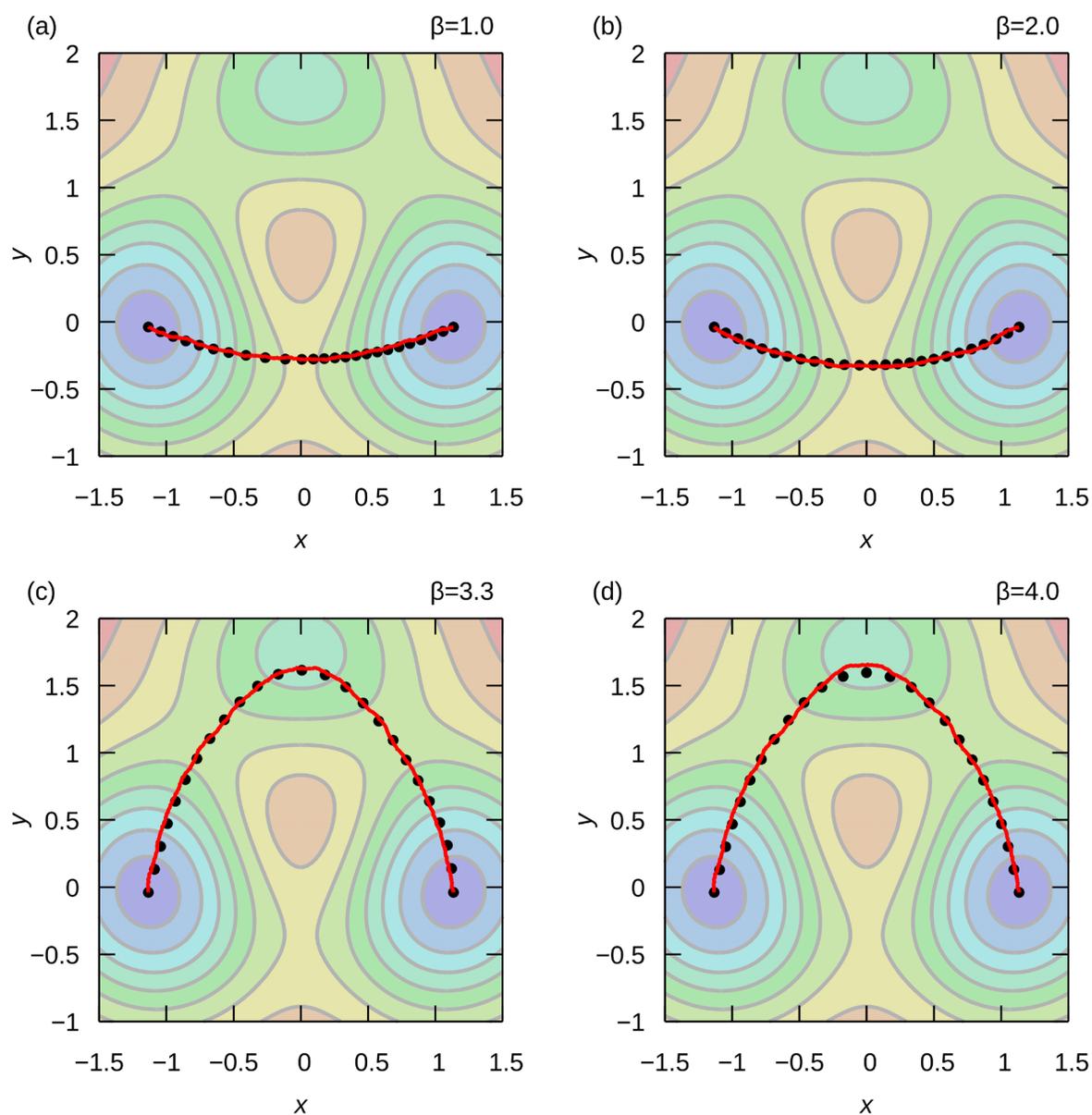

**Fig. S2.** Comparison of paths obtained by the present method (red lines), and MaxFlux method (nodes are shown by black circles) for $\beta =$ (a) 1.0, (b) 2.0, (c) 3.3, and (d) 4.0. For the present method, the grid density is set to $\rho_{\text{grids}} =$ 40000.